\documentclass[aps,prd,showpacs,preprintnumbers,nofootinbib,twocolumn]{revtex4}
\usepackage[dvips]{graphicx}
\usepackage{bm,latexsym,amsmath,amssymb,amsfonts}
\newcommand*{\cT}{{\cal T}}
\newcommand*{\cS}{{\cal S}}
\newcommand*{\cA}{{\cal A}_{j\ell mn}}
\newcommand*{\rin}{{\rm in}}
\newcommand*{\rout}{{\rm out}}
\newcommand*{\D}{{\rm d}}
\begin{document}

\title{Bulk scalar emission from a rotating black hole pierced by a tense brane}

\author{Tsutomu~Kobayashi}
\email[Email: ]{tsutomu"at"gravity.phys.waseda.ac.jp}
\author{Masato~Nozawa}
\email[Email: ]{nozawa"at"gravity.phys.waseda.ac.jp}
\author{Yu-ichi Takamizu}
\email[Email: ]{takamizu"at"gravity.phys.waseda.ac.jp}
\affiliation{Department of Physics, Waseda University, Okubo 3-4-1, Shinjuku, Tokyo 169-8555, Japan}

\begin{abstract}
We study the emission of scalar fields into the bulk from a six-dimensional
rotating black hole pierced by a 3-brane.
We determine the angular eigenvalues in the presence of finite brane tension
by using the continued fraction method.
The radial equation is integrated numerically, giving the absorption probability (graybody factor)
in a wider frequency range than in the preexisting literature.
We then compute the power and angular momentum emission spectra
for different values of the rotation parameter and brane tension,
and compare their relative behavior in detail.
As is expected from the earlier result for a nonrotating black hole,
the finite brane tension suppresses the emission rates.
As the rotation parameter increases, the power spectra are reduced at low frequencies
due to the smaller Hawking temperature and are
enhanced at high frequencies due to superradiance.
The angular momentum spectra are enhanced over the whole frequency range
as the rotation parameter increases.
The spectra and the amounts of energy and angular momentum radiated away
into the bulk are thus determined by the interplay of these effects.
\end{abstract}

\preprint{WU-AP/277/07}
\maketitle

\section{Introduction}

Braneworld models with large extra dimensions~\cite{ADD, koko, RS}
bring us an interesting possibility to address the hierarchy problem
by lowering the fundamental scale of gravity
down to order of TeV.
It has been argued in the context of TeV scale gravity that
mini black holes might be created
through high-energy particle collision at future colliders~\cite{LHC}.
Much effort has been directed towards a theoretical understanding
of the black hole formation at TeV energies (e.g.,~\cite{collision}).
After their production, the black holes will decay via Hawking radiation~\cite{Hawking}.
This process provides a window to probe high-energy physics,
gravity at small distances, and properties of extra dimensions,
which motivates recent extensive studies on this topic.
A nonexhaustive sampling of the literature can be found in Refs.~\cite{EHM2,KantiMarch,
Harris,Duffy,HarrisKanti,BraneDecay2,Creek,Creek2,Berti:2003yr,IUM,Morisawa:2004fs,Shijun,Vitor,
Kanti:2005xa,IOP1,IOP2,IOP3,FS,Kanti,Alan,Wade}.
For a review see Ref.~\cite{Kanti:review}.
Most of the related work to date has ignored the effect of brane tension,
treating black holes as ``isolated'' ones (see, however, Refs.~\cite{Fro, Gin, BlackMax, nino}
for the effects of self-gravity of branes).

It is in general very difficult to obtain a black hole solution localized
on a brane with finite tension because tension
curves the brane as well as the bulk (cf. Refs.~\cite{EHM, KTN, stars, Tanahashi}).
However, codimension-2 branes exceptionally allow for
a simple construction of localized black holes
thanks to their special property; starting from
the Myers-Perry solution~\cite{MP}
one rescales the polar angle around a symmetry axis as $\psi\to B\psi$ and then
the brane tension is proportional to the deficit angle $2\pi(1-B)$.
In this way both nonrotating~\cite{KaloperKiley} and rotating~\cite{Kiley}
black holes on codimension-2 branes have been constructed.
Following the work of~\cite{KaloperKiley}, Hawking evaporation~\cite{Dai} and
the quasi-normal modes for bulk scalars~\cite{Chen, alBinni} and fermions~\cite{Cho}
have been investigated in the nonrotating background,
showing that the finite brane tension modifies the standard result
derived assuming negligible tension.

In this paper, we shall
consider a six-dimensional {\em rotating} black hole pierced by a tense 3-brane
and discuss the emission of massless scalar fields into the bulk.
We intend to shed light on the spin-down phase in the life of a black hole,
which is often neglected in the literature but could be of some significance.
(In fact, a rotating black hole does not necessarily spin-down to zero, but
evolves toward a nonzero angular momentum~\cite{Spinning-down, Nomura}.)
Ignoring the brane tension, very recently Creek {\em et al.} 
studied the emission of scalars in the bulk
in a higher-dimensional rotating black hole background~\cite{Kanti}.
They employed matching techniques to obtain an analytic solution to the scalar field equation,
which is a good approximation in the low-energy ($\omega r_h\ll 1$) and
slow-rotation ($a/r_h\ll 1$) regime, where $\omega$ is the energy of the emitted particle,
$r_h$ is the black hole horizon radius, and $a$ is the rotation parameter.
In the present paper,
with the help of numerical computations we are able to handle the intermediate regime
($\omega r_h\gtrsim 1$ and $a/r_h\gtrsim 1$), and thus
we not only include the effect of the finite tension but also
extend the range of validity of~\cite{Kanti}.

This paper is organized as follows. In the next section we give a quick review of
the rotating black hole solution on a codimension-2 brane.
In Sec.~III we present separated equations of motion for a massless scalar field
and determine angular eigenvalues in the presence of the deficit angle.
Then in Sec.~IV the radial equation is solved numerically to give the power and
angular momentum emission spectra. Finally we summarize our conclusions in Sec.~V.
Appendix contains the analytic calculation of the absorption probability,
which complements the numerical results presented in the main text.

\section{A rotating black hole on a codimension two brane}

We begin with a brief review
of the rotating black hole solution on a codimension-2 brane.
(For further detail see Refs.~\cite{KaloperKiley, Kiley}.)
The solution shares some properties with the Myers-Perry black
hole \cite{MP}.
We are considering the models with five spatial dimensions,
and so the rotation group
is $SO(5)$. The number of Casimirs (i.e., the number of mutually commuting  
elements of the group) is equal to rank$[SO(5)]=2$. Hence, we have 
two axes of rotation associated with two angular momenta.
However, in the present article 
we will be focusing on the special but simple case of a single rotation parameter
with the angular momentum pointing along the brane.
This is indeed an interesting case from the phenomenological point of
view, because the black hole formed by the collision of
two particles confined to the brane will have a single rotation parameter.
The exact metric that describes such a rotating black hole
is given by~\cite{Kiley}
\begin{align}
\D s^2 =& -\left(1-\frac{\hat\mu}{ r\rho^2}\right)\D t^2
-\frac{2\hat\mu a}{r\rho^2}\sin^2\theta
\D t\D \phi+\frac{\rho^2}{\Delta}\D r^2
\nonumber\\&
+\rho^2\D \theta^2+\sin^2\theta\left(r^2+a^2+\frac{\hat\mu
 a^2}{r\rho^2}\sin^2\theta\right)\D \phi^2
\nonumber\\&
+r^2\cos^2\theta\left(\D 
\chi^2+B^2\sin^2\chi \D \psi^2\right),
\label{BH}
\end{align}
where
\begin{eqnarray}
\Delta:=r^2+a^2-\frac{\hat\mu}{r},\quad\rho^2:=r^2+a^2\cos^2\theta.
\end{eqnarray}
The coordinate ranges are $0\leq \theta \leq \pi /2, \,0\leq \chi \leq \pi$, and
$0\leq \phi, \psi <2\pi$. 
The parameter $B$ is related to the brane tension $\sigma$ as
\begin{eqnarray}
B=1-\frac{\sigma}{2\pi M_*^4},
\end{eqnarray}
where $M_*$ is the six-dimensional fundamental scale.
We assume that $0<B\leq1$.
When $B=1$, the above metric reduces to the usual Myers-Perry solution
with a single rotation parameter in six dimensions~\cite{MP}. 
When $B\neq 1$ the solution is asymptotically conical.
The parameters $\hat \mu \,(>0) $ and $a$ denote the specific mass and angular momentum,
respectively,
related to the ADM mass and angular momentum of the black hole as
\begin{eqnarray}
M_{{\rm BH}}=2A_4 M_*^4\,\hat\mu\, B,
\quad J_{{\rm BH}}=\frac{1}{2}M_{{\rm BH}} a,
\label{MandJ}
\end{eqnarray}
where $A_d=2\pi^{(d+1)/2}/\Gamma[(d+1)/2]$
is the area of a unit $d$-sphere.
Note here that the effect of the deficit angle $B$ is separated out from
the definition of the area.
The black hole horizon radius $r_h$ follows from $\Delta(r_h)=0$.
For later purpose
it is convenient to define the dimensionless measure of the angular
momentum $a_*:=a/r_h$. Since the sign flip $a\to -a$ simply changes the
direction of rotation, in what follows we will assume $a\ge 0$ 
without any loss of generality.

Note that $\Delta (r_h)=0$ has a root
for arbitrary $a$. This should be contrasted with
the four-dimensional Kerr black hole, which has an upper bound on the
rotation parameter. This point will become clearer by considering the black hole temperature.
The generator of the horizon is parallel to the Killing vector 
$\xi^\mu =(\partial /\partial t)^\mu+\Omega _h (\partial /\partial \phi)^\mu$,
and the surface gravity of the black hole is given by
$\kappa^2 =-(\nabla^\mu \xi^\nu)(\nabla_\mu\xi_\nu)/2$.
Here $\Omega_h$ is the angular velocity of the horizon:
\begin{align}
\Omega_h:=\frac{a}{r_h^2+a^2}.
\end{align}
Then, the Hawking temperature is given by~\cite{Hawking}
\begin{eqnarray}
T_H:=\frac \kappa {2\pi}=\frac{3+a_*^2}{4\pi(1+a_*^2)r_h},
\label{temp}
\end{eqnarray}
and one sees that the temperature
never attains zero. This means the black hole is manifestly nondegenerate
for any value of $a$.

An interesting effect of the brane tension perhaps one is immediately aware of
is the rescaling of the gravitational scale $M_* \to M_{*{\rm eff}}=B^{1/4}M_*$,
as is observed in Eq.~(\ref{MandJ}).
This modifies the scenario of the black hole production and evaporation
as follows~\cite{KaloperKiley, Kiley, Dai}.
Suppose that a black hole is formed through a particle collision and
its mass $M_{{\rm BH}}$ is fixed by the accelerator.
The upper limit on the impact parameter $b$ is given by $b\leq 2r_h$.
Since $r_h \sim M^{1/3}_{{\rm BH}}/M_{*{\rm eff}}^{4/3}$,
for fixed $M_{{\rm BH}}$ the horizon radius is {\em larger} than the naive estimate,
$r_h \sim B^{-1/3}$, which is called the ``lightning rod'' effect~\cite{KaloperKiley,Kiley}.
In the present situation producing black holes will become easier due to this effect.
Since the black hole formed in this way has the angular momentum $J=bM_{{\rm BH}}/2$,
we have $J\leq J_{{\rm max}}:=M_{{\rm BH}}r_h$,
and so the angular momentum can also be {\em larger} than naively expected,
$J_{{\rm max}}\sim B^{-1/3}$.
Now the time scale $\tau$ of mass loss is estimated from
$\D M_{{\rm BH}}/\D t\sim$ (horizon area)$\times T_H^6\sim Br_h^{-2}$,
yielding $\tau\sim B^{-5/3}$.
Thus the lifetime of the black hole will become longer.
Similarly, the time scale $\tau'$ of angular momentum loss is estimated roughly from
$\D J_{{\rm BH}}/\D t\sim r_h \D M_{{\rm BH}}/\D t$. This gives $\tau'\sim B^{-5/3}$,
and hence the time scale $\tau'$ will also become larger.

The above mentioned modifications are caused by the rescaling of the gravitational scale.
In the rest of the paper 
we separate this effect and
discuss
the impact of finite brane tension on Hawking emission
{\em with the horizon radius $r_h$ fixed}.

Let us finally remark that
in models with extra dimensions the bulk will be compact.
Throughout the paper we assume that the horizon size of the black hole
is smaller than the typical compactification scale.
With this, the above solution provides
a nice description of a six-dimensional black hole localized on a 3-brane.

\section{Scalar wave equation in a rotating black hole background}

We now turn to discuss Hawking emission of massless scalar fields.
To this end we solve the equation of motion for a massless scalar
field $\Psi$ in the rotating black hole background~(\ref{BH}). 
The governing equation is given by
\begin{eqnarray}
\frac{1}{\sqrt{-g}}\partial_{\mu}\left(\sqrt{-g}\;g^{\mu\nu}\partial_{\nu}\Psi\right)=0,
\end{eqnarray}
where $\sqrt{-g}=B\rho^2r^2\sin\theta\cos^2\theta\sin\chi$.
We can separate the above equation by assuming the ansatz
\begin{eqnarray}
\Psi=e^{-i\omega t}e^{im\phi}R(r)S(\theta)T(\chi)\Xi(\psi),
\end{eqnarray}
where $m=0, \pm1, \pm2, ...$
The radial equation is
\begin{eqnarray}
\frac{1}{r^2}\frac{\D }{\D r}\!
\left(r^2\Delta \frac{\D }{\D r}R\right)
+\left[
\frac{K^2}{\Delta}-\nu(\nu+1)\frac{a^2}{r^2}
-\lambda\right]R=0,
\label{radial_eq}
\end{eqnarray}
where
\begin{eqnarray}
K:=(r^2+a^2)\omega-am,\quad\lambda:=\eta+a^2\omega^2-2am\omega,
\end{eqnarray}
and the angular equations are given by
\begin{align}
\frac{1}{\sin\theta\cos^2\theta}\frac{\D }{\D \theta}
\left(\sin\theta\cos^2\theta\frac{\D }{\D \theta}S\right)
\qquad \qquad \qquad&
\nonumber\\
+\left[a^2\omega^2\cos^2\theta-\frac{m^2}{\sin^2\theta}-\frac{\nu(\nu+1)}{\cos^2\theta}\right]S
=-\eta S,&&\label{ang_S}
\\
\frac{1}{\sin\chi}\frac{\D }{\D \chi}\left(\sin\chi\frac{\D }{\D\chi}T\right)
-\frac{\mu^2}{\sin^2\chi}T=-\nu(\nu+1)T,&
\\
\frac{1}{B^2}\frac{\D ^2}{\D \psi^2}\Xi=-\mu^2\Xi.\label{eqXI}&
\end{align}
Here $\eta, \nu,$ and $\mu$ are separation constants.
The above equations of motion are the same as those derived in Ref.~\cite{IUM} except
that in the last equation~(\ref{eqXI}) there appears ``$B$.''
We immediately see that
\begin{eqnarray}
\mu=\frac{n}{B}\quad\text{with}\quad n=0,\pm1, \pm2, ...
\end{eqnarray}
As is emphasized in Ref.~\cite{Dai}, the eigenvalues are coupled and therefore
all of the eigenvalues will be modified when $B\neq 1$.
Since the eigenvalues $\nu$ and $\eta$ will be dependent on $|n|$,
we can restrict to $n\geq 0$ without any loss of generality.
Negative values of $n$ can be treated similarly.


\begin{figure}[tb]
  \begin{center}
    \includegraphics[keepaspectratio=true,height=50mm]{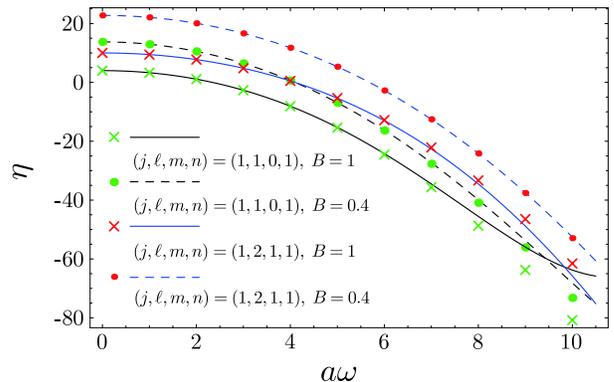}
  \end{center}
  \caption{Eigenvalue $\eta$ for selected values of $(j, \ell, m, n)$ and $B$.
  Solid and dashed lines refer to
  the series expansion~(\ref{eingen_expansion}) truncated at 7th order, while
  points and crosses show the numerical results without relying on the small-$a\omega$ expansion.}%
  \label{fig:eigenvalue.eps}
\end{figure}

We proceed to determine the eigenvalue $\nu$.
Performing the change of the variable and function as
$x=\cos\chi$, $\cT:=(1-x^2)^{-\mu/2}T$,
we have
\begin{eqnarray}
&&(1-x^2)\frac{\D ^2}{\D x^2}\cT
-2\left(\mu+1\right)x\frac{\D }{\D x}\cT
\nonumber\\&&\qquad\qquad\qquad
+(\nu-\mu)(\nu+\mu+1)\cT=0.
\end{eqnarray}
The boundary conditions are given by
\begin{eqnarray}
\frac{\D }{\D x}\cT=\pm\frac{(\nu-\mu)(\nu+\mu+1)}{2(\mu+1)}\cT
\quad \text{at} \quad x=\pm 1.
\end{eqnarray}

As in~\cite{alBinni}, we first consider the special case in which $B=1/N$ with $N=1, 2, ...$
and hence $\mu=nN$ is an integer.
In this case it is easy to find $\nu=\mu, \mu+1, \mu+2, ...$
Thus the eigenvalue $\nu$ can be written in terms of an integer $j$ as
\begin{eqnarray}
\nu = j+b_n,\quad b_n:=\frac{1-B}{B}|n|,
\label{eigen_nu}
\end{eqnarray}
where $j\geq |n|$ and the absolute value signs are inserted for clarity.
Although Eq.~(\ref{eigen_nu}) was derived for $B^{-1}=1, 2, ...$,
this result can be generalized to arbitrary $B\in(0,1]$~\cite{alBinni}.
We have confirmed numerically that Eq.~(\ref{eigen_nu}) indeed holds
for noninteger values of $B^{-1}$.

When $B=1$, $\nu$ is independent of $n$ (and as a result
$\eta$ is also independent of $n$).
For the tensional case, however, $\nu$ depends on $n$ (and hence does $\eta$).
From Eq.~(\ref{eigen_nu})
we see that the brane tension {\em increases} the eigenvalue $\nu$
relative to the tensionless case.

To determine the eigenvalue $\eta$,
we exploit the continued fraction method
developed originally by Leaver~\cite{Leaver}. 
Following~\cite{Cardoso}, we write $S=(1-u^2)^{|m|/2}u^{\nu}\cS$ with $u=\cos\theta$.
Then, $\cS$ obeys
\begin{eqnarray}
(1-u^2)\frac{\D^2}{\D u^2}\cS+2\left[(\nu+1)\frac{1-u^2}{u}-(|m|+1)u\right]\frac{\D}{\D u}\cS
\nonumber\\
+\left[
a^2\omega^2u^2+\eta-\nu(\nu+3)-|m|(|m|+2\nu+3)
\right]\cS=0,
\nonumber\\\label{calSeq}
\end{eqnarray}
subject to the regularity boundary conditions
\begin{eqnarray}
\frac{\D}{\D u}\cS=0
\qquad\qquad\text{at}\quad u=0,\qquad&&
\\
\frac{\D}{\D u}\cS=
\frac{a^2\omega^2+\eta-\nu(\nu+3)-|m|(|m|+2\nu+3)}{|m|+1}\cS&&
\nonumber\\
 \text{at}\quad u=1.\qquad&&
\end{eqnarray}

We seek for a series solution in the form of
\begin{eqnarray}
\cS=\sum_{p=0}^{\infty}a_pu^{2p}.\label{ssol}
\end{eqnarray}
Substituting this to Eq.~(\ref{calSeq}) we obtain the three-term recursion relation
\begin{eqnarray}
\alpha_0 a_1+\beta_0 a_0&=&0,
\\
\alpha_pa_{p+1}+\beta_pa_p+\gamma_pa_{p-1}&=&0,\quad p=1, 2, ...,
\end{eqnarray}
where
\begin{eqnarray}
\alpha_p&:=&2(p+1)(2\nu+2p+3),
\\
\beta_p&:=&\eta-(\nu+|m|+2p)(\nu+|m|+2p+3),
\\
\gamma_p&:=&(a\omega)^2.
\end{eqnarray}
The continued fraction equation for the eigenvalue is then given by
\begin{eqnarray}
\beta_0-\cfrac{\alpha_0\gamma_1}{\beta_1-\cfrac{\alpha_1\gamma_2}{\beta_2-
\cfrac{\alpha_2\gamma_3}{\beta_3-\cdots}}}=0.\label{cfe}
\end{eqnarray}
The boundary conditions are automatically satisfied by the series solution~(\ref{ssol}).

We expand the eigenvalue $\eta$ in powers of $a\omega$ around $a\omega=0$:
\begin{eqnarray}
\eta=\sum_{p=0}^{\infty}\tilde\eta_p(a\omega)^p.
\label{eingen_expansion}
\end{eqnarray}
In order for the series to converge in the limit $a\omega\to0$, we require
that it has a finite number of terms. Imposing $\beta_q=0$ for some integer $q\geq0$ and
identifying $2q=\ell-(j+|m|)$,
we obtain
\begin{eqnarray}
\eta = (\ell+b_n)(\ell+3+b_n)
\end{eqnarray}
for $a\omega=0$.
Here we have a restriction for the integer $\ell$:
\begin{eqnarray}
\frac{1}{2}(\ell-j-|m|)=0, 1, 2, ...
\end{eqnarray}
To determine the
coefficients in~(\ref{eingen_expansion}) it is convenient to use the $q$th inversion of~(\ref{cfe}):
\begin{eqnarray}
\beta_q-\cfrac{\alpha_{q-1}\gamma_q}{\beta_{q-1}-\cfrac{\alpha_{q-2}\gamma_{q-1}}{\beta_{q-2}-\cdots}}
=\cfrac{\alpha_q\gamma_{q+1}}{\beta_{q+1}-\cfrac{\alpha_{q+1}\gamma_{q+2}}{\beta_{q+2}-\cdots}}.
\end{eqnarray}
With some manipulation
we find $\tilde\eta_1=\tilde\eta_3=0$ and
\begin{widetext}
\begin{eqnarray}
\tilde\eta_0&=& (\ell+b_n)(\ell+3+b_n),
\label{et0}\\
\tilde\eta_2&=&-\frac{2\ell(\ell+3)+3+2j+2j^2-2m^2+4(\ell+j+2)b_n+4b_n^2}{[2(\ell+b_n)+1][2(\ell+b_n)+5]},
\label{et2}
\\
\tilde\eta_4&=&\frac{(\ell-j-|m|)(\ell+j-|m|+1+2b_n)}{4[2(\ell+b_n)+1]^2}
\left[\frac{(\ell-j-|m|-2)(\ell+j-|m|-1+2b_n)}{4[2(\ell+b_n)-1]}-\tilde\eta_2\right]
\nonumber\\&&
-\frac{(\ell-j-|m|+2)(\ell+j-|m|+3+2b_n)}{4[2(\ell+b_n)+5]^2}
\left[
\frac{(\ell-j-|m|+4)(j+\ell-|m|+5+2b_n)}{4[2(\ell+b_n)+7]}
+\tilde\eta_2
\right].\label{et4}
\end{eqnarray}
\end{widetext}
One can confirm that Eqs.~(\ref{et0})--(\ref{et4}) correctly reproduce
the known result in the case of $b_n=0$~\cite{Cardoso, Kanti}.\footnote{The second order
coefficient in~\cite{Cardoso} contains a sign error, which has been corrected in~\cite{Kanti}.}
Although the expressions are too lengthy,
higher order coefficients can be obtained easily.

One can instead integrate Eq.~(\ref{ang_S}) numerically to determine $\eta$.
The numerical computation has an advantage that it does not rely on
the small-$a\omega$ expansion.
In Fig.~\ref{fig:eigenvalue.eps}
we compare the series expansion truncated at 7th order
with the numerical result that is free from any approximation. It can be seen from this that
the analytic result can reproduce the numerical computation
remarkably well even for $a\omega\sim 6$.
Therefore, in the next section we will safely use the analytic approximation for $\eta$.
We thus avoid numerical determination of the eigenvalues and
so the rest of the problem simply reduces
to solving the ordinary differential equation for the radial mode function.

We remark here that while the brane tension does not affect the $n=0$ mode,
the eigenvalue $\eta$ (for $n\neq0$) increases as $B$ decreases (and so
the tension increases), as can be seen from Fig.~\ref{fig:eigenvalue.eps}.
This behavior has been observed for a nonrotating black hole in~\cite{Dai},
and now it turns out that the similar thing generally holds for a rotating one.

\section{Power and angular momentum emission spectra}

\begin{figure}[b]
  \begin{center}
    \includegraphics[keepaspectratio=true,height=45mm]{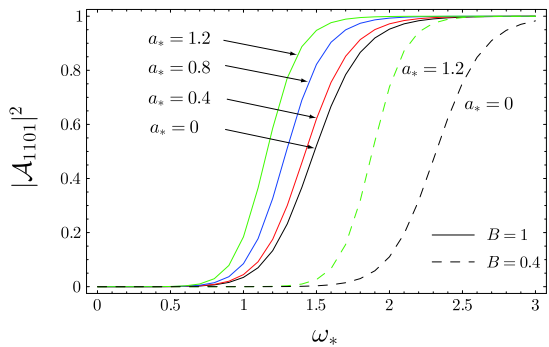}
  \end{center}
  \caption{Absorption probabilities for $(j, \ell, m, n)=(1,1,0,1)$, $B=1, 0.4$, and various $a_*$.}%
  \label{fig:ab1.eps}
\end{figure}

\begin{figure}[tb]
  \begin{center}
    \includegraphics[keepaspectratio=true,height=45mm]{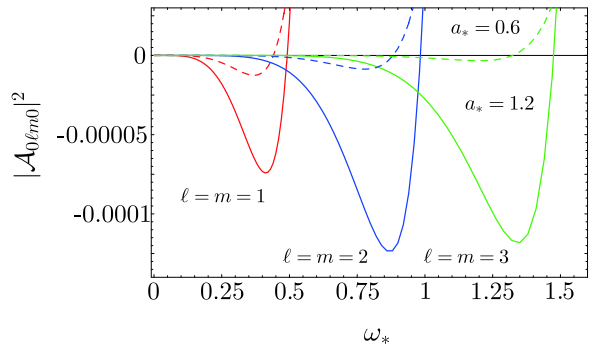}
  \end{center}
  \caption{Superradiance modes for $a_*=1.2$ and $0.6$. The
  absorption probabilities are negative for $\omega_*<m\Omega_hr_h$, where
  $\Omega_hr_h\simeq 0.49$ for $a_*=1.2$ and $\Omega_hr_h\simeq0.44$ for $a_*=0.6$.}%
  \label{fig:sr.eps}
\end{figure}

We are going to solve the radial equation~(\ref{radial_eq}) to compute
the emission spectra.
To do so we first specify the asymptotic form of the solution 
close to the horizon and far away from it.
In terms of the new coordinate defined by $\D r_*:=[(r^2+a^2)/\Delta]\D r$
and the function $\Phi:=r(r^2+a^2)^{1/2}R$, Eq.~(\ref{radial_eq}) can be rewritten as
\begin{eqnarray}
\frac{\D ^2}{\D r_*^2}\Phi+\left\{[\omega-m\Omega(r)]^2-V(r)\right\}\Phi=0.
\end{eqnarray}
where $\Omega(r):=a/(r^2+a^2)$ and
\begin{align}
V(r):=&\frac{\Delta}{(r^2+a^2)^2}\left[\lambda+\nu(\nu+1)\frac{a^2}{r^2}\right]
\nonumber\\&
+\frac{1}{r(r^2+a^2)^{1/2}}\frac{\D ^2}{\D r_*^2}
\left[r(r^2+a^2)^{1/2}\right].
\label{pot_V}
\end{align}
It is easy to see that $V(r)\approx 0$ as $r\to r_h$ and $r\to\infty$.
Keeping this in mind the asymptotic solutions are found to be
\begin{eqnarray}
r\to r_h:&&R\simeq A^{(h)}_\rin e^{-iky}+A^{(h)}_\rout e^{iky},
\\
r\to \infty:&& R\simeq A^{(\infty)}_\rin\frac{e^{-i\omega r}}{r^2}
+A^{(\infty)}_\rout\frac{e^{i\omega r}}{r^2},
\end{eqnarray}
where
\begin{eqnarray}
k:=\omega-m\Omega_h,
\label{k-Omega_h}
\end{eqnarray}
and
$y$ is the tortoise-like coordinate defined by 
\begin{eqnarray}
y:=\frac{1+a_*^2}{3+a_*^2}\,r_h\ln\left[\frac{\Delta(r)}{r^2+a^2}\right].
\label{tortoise}
\end{eqnarray}
We choose the boundary condition $A^{(h)}_{\rout}=0$, i.e.,
we impose that no outgoing modes exist near the horizon.
The absorption probability is then given by
\begin{eqnarray}
|\cA|^2 = 1-\left|\frac{A^{(\infty)}_\rout}{A^{(\infty)}_\rin}\right|^2.
\end{eqnarray}
Note the explicit dependence on the angular eigenvalue $n$ of the absorption probability.
This is due to the nonzero brane tension $(B\neq 1)$.

Using the series expansion of the eigenvalue $\eta$
(truncated at 7th order), we numerically integrate Eq.~(\ref{radial_eq})
and compute the absorption probability.
We defer to Appendix the analytic calculation of 
the absorption probability
in the limit $\omega _*:=\omega r_h\ll 1$ and $a_*\ll 1$.
The analytic and numerical results are found to be in good agreement
in the regime $\omega _*\ll 1$ and $a_*\ll 1$.

As was remarked above, both $\eta$ and $\nu$ become larger
as the brane tension increases. This results in the enhancement
of the ``potential''~(\ref{pot_V}), which will reduce the absorption probability.
A typical example of the absorption probability (as a function of $\omega_*$)
is plotted for various $a_*$ and $B$
in Fig.~\ref{fig:ab1.eps},
showing that
the absorption probability indeed decreases with decreasing $B$.
We can see in Fig.~\ref{fig:sr.eps} that there appear superradiant modes
with $0<\omega<m\Omega_h$, for which the absorption probability is negative~\cite{superr}.

\begin{figure}[tb]
  \begin{center}
    \includegraphics[keepaspectratio=true,height=50mm]{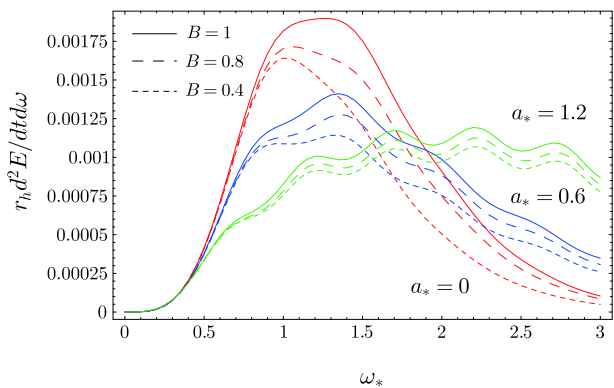}
  \end{center}
  \caption{Power emission spectra for different values of $a_*$ and $B$.}
  \label{fig:E_sp.eps}
\end{figure}

\begin{figure}[tb]
  \begin{center}
    \includegraphics[keepaspectratio=true,height=50mm]{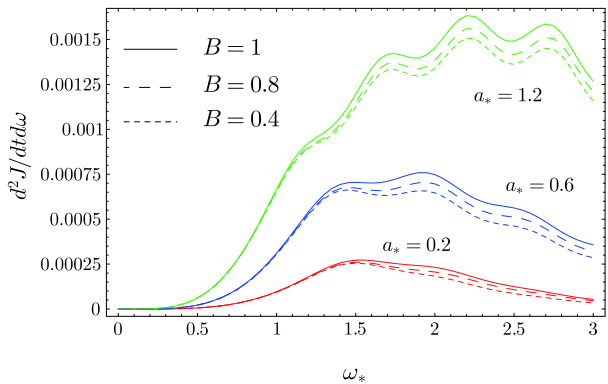}
  \end{center}
  \caption{Angular momentum spectra for different values of $a_*$ and $B$.}%
  \label{fig:J_sp.eps}
\end{figure}

From the absorption probability we compute the energy and angular momentum
emission rates. They are given by the formulas
\begin{eqnarray}
\frac{\D ^2\! E}{\D t\D \omega}&=&\frac{1}{2\pi}\sum_{j, \ell, m, n}\frac{\omega}{e^{k/T_H}-1}
\left|\cA\right|^2,
\\
\frac{\D ^2\! J}{\D t\D \omega}&=&\frac{1}{2\pi}\sum_{j, \ell, m, n}\frac{m}{e^{k/T_H}-1}
\left|\cA\right|^2,
\end{eqnarray}
where $T_H$ and $k$ were already defined in Eqs.~(\ref{temp}) and~(\ref{k-Omega_h}), respectively.
We summed up to $\ell=5$ modes in calculating the above quantities.
Our numerical results are summarized in Figs.~\ref{fig:E_sp.eps} and~\ref{fig:J_sp.eps}.

\begin{figure}[tb]
  \begin{center}
    \includegraphics[keepaspectratio=true,height=50mm]{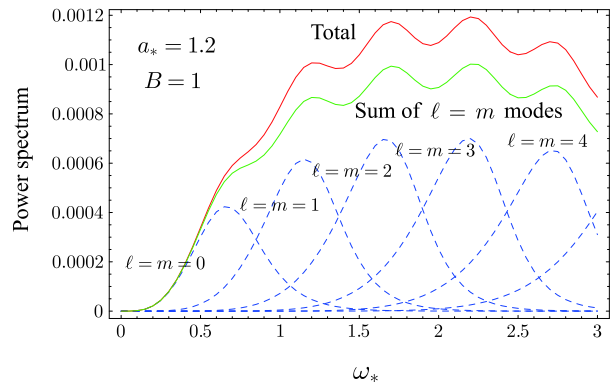}
  \end{center}
  \caption{Contributions from $\ell=m$ modes. Red (solid) line shows the total power emission
  spectrum depicted also in Fig.~\ref{fig:E_sp.eps}, while green (solid) line represents the sum of $\ell=m$
  modes, each of which is shown by blue (dashed) line.}%
  \label{fig:lm.eps}
\end{figure}

We find that the finite brane tension reduces the power and angular momentum emission
spectra. This is anticipated from the behavior of the absorption probability stated above.
The power emission rates are reduced in the low frequency regime
as the rotation parameter $a_*$ increases.
This is because the Hawking temperature becomes lower with increasing $a_*$.
However, the power emission rates are enhanced at high frequencies.
This should be caused by superradiance.
The angular momentum emission rates are enhanced over the whole frequency range
with increasing $a_*$
because the effects of superradiance win out.
(In the previous estimates~\cite{Spinning-down,Nomura,Nozawa}
significant superradiance is observed
as the angular momentum increases.)
One can see from Fig.~\ref{fig:lm.eps}
that a large portion of the contribution to the power emission spectrum is coming from
the $\ell=m$ modes, which show the superradiant behavior (except for $\ell=m=0$).
We confirmed that the same is true for the angular momentum spectrum.
As $a_*$ increases the amplitudes of these modes
are enhanced, lifting up the total spectra.
One can also see that the oscillatory behavior of the spectra is due to these modes,
with each peak corresponding to each mode.
However, the above things do not mean that only the $\ell = m$ modes are important
in determining the behavior of the spectra;
the eigenvalues for the $\ell = m$ modes are independent of the conical deficit and
the other modes ($|n|\geq 1$) determine the decrease in the spectra with decreasing $B$.

Since the brane induced metric is the same as the tensionless case,
the Hawking spectrum of brane-localized fields
does not depend on the tension (for fixed $r_h$).\footnote{Although
this statement is true in the present model, it is not clear whether or not the
induced metric is independent of the brane tension in more realistic situations with compact
extra dimensions.}
The emission of brane scalar fields
has been studied in Refs.~\cite{IOP1, IOP2, HarrisKanti, Duffy, Creek}
for rotating black holes in various dimensions,
and the brane-to-bulk ratio of the energy emission rates has been
discussed in~\cite{Kanti},
implying that the dominant channels are the brane-localized modes.
Since the finite brane tension further suppresses the bulk field contributions,
it is likely that brane-localized scalar emission dominates
bulk emission also in the present case.
Although bulk modes will not be observed directly,
bulk emission is still important because it
indirectly determines the amount of energy and angular momentum
left for brane-localized emission.

\section{Concluding remarks}

The black hole production in
TeV scale gravity offers us a possible window to explore the presence
of extra dimensions.
In this paper we have studied Hawking emission of scalar fields into the bulk
from a rotating black hole localized on a codimension-2 brane.
The exact solution we used is a rare example in which
we can treat self-gravity of the brane consistently
in the context of brane-localized black holes~\cite{Kiley}.
This simple model enables us to elucidate how the brane tension
modifies the Hawking spectra relative to the tensionless case.

Assuming the separable ansatz for the scalar field,
we have determined the angular eigenvalue in the series expansion form:
$\eta=\tilde\eta_0+\tilde\eta_2(a\omega)^2+\cdots$.
This analytic approximation (truncated at 7th order) was turned out to be in excellent agreement with
the numerical result in the regime $a\omega\lesssim 6$.
Using the analytic form of $\eta$, we then integrated the radial equation numerically
and computed the power and angular momentum emission spectra.
Our finding is that the finite brane tension suppresses
the power and angular momentum spectra.
We also showed that the power emission rates are
reduced at low frequencies and enhanced at high frequencies
as the rotation parameter $a_*$ increases.
The suppression at low frequencies is due to the smaller Hawking temperature
and the enhancement at high frequencies is caused by superradiance.
The angular momentum emission rates are enhanced over the whole frequency range
as $a_*$ increases due to superradiance.
The spectra and the amounts of energy and angular momentum
radiated away into the bulk are thus determined by the interplay of these effects.
To conclude, the brane tension plays an important role in
the evaporation process in the life of a mini black hole.

It is possible to obtain an analytic but approximate solution to the radial equation,
as has been done recently in~\cite{Kanti} and is replicated in Appendix.
The range of validity of this analytic approximation is restricted to $\omega_*\ll 1$ and $a_*\ll 1$.
Therefore, even in the tensionless case ($B=1$) our result is new,
in that we have extended the range of validity of~\cite{Kanti}
by invoking the numerical approach.
We however exploited the analytic expression for $\eta$ to
simplify numerical calculations.
The eigenvalues $\eta$ obtained here by the continued fraction method
are quite accurate even for $a_*\omega_*\sim 6$,
which allows us to compute the spectra for $\omega_*\gtrsim 1$ and $a_*\gtrsim 1$.
In order to explore the full regime extending to $a\omega \gtrsim {\cal O}(10)$,
we need to solve both the radial and angular equations numerically,
which is left to further investigation.

In this paper we considered only the emission of scalar fields.
It would be interesting to study the bulk emission of higher spin fields
and determine the brane-to-bulk ratio of the energy and angular momentum emission rates.
Recently, it has been reported that for fermion fields
the bulk emission dominates the brane-localized emission in a six- or higher dimensional
Schwarzschild background~\cite{Wade}.
Therefore, investigating the effects of the finite brane tension~\cite{Cho} and the black hole rotation
on the fermion emission
would be of particular interest.
Another open issue is to clarify the spin-down evolution of the rotating black hole
and the effect of brane tension on it.
This process can be studied along the line of~\cite{Spinning-down,Nomura}.
We plan to return to this issue in the near future.
Finally, it is fair to say that all of the results in this paper have been derived
ignoring the compactification mechanism, which may affect both brane and bulk emissions.
This point is worth exploring, though constructing brane-localized black hole solutions
in compact space will be quite difficult.


\acknowledgments
We would like to thank Sam Dolan and Marc Casals for useful comments.
TK, MN, and YT are supported by the JSPS under Contract Nos.~19-4199, 
19-204, and 17-53192.


\appendix

\section{Analytic approximation method for $\omega_*\ll1$ and $a_*\ll1$}

In this appendix we present an approximation method to obtain
an analytic expression for the absorption probability.
Our result here simply generalizes that of~\cite{Kanti} to include the
deficit angle (see also \cite{Creek}). The procedure is as follows:
first we obtain the asymptotic solutions in the far-field and near-horizon regions,
and then match the two solutions in the intermediate zone.
The approximation is valid in the low-energy ($\omega_*\ll1$) and
slow-rotation ($a_*\ll1$) regime.
We check that
in this range the analytic result agrees
with the numerical one displayed in the main text.

Let us first focus on the near-horizon zone ($r\simeq r_h$). It is convenient to work with
a new radial variable defined by
\begin{eqnarray}
\xi:=\frac{\Delta}{r^2+a^2},
\end{eqnarray}
for which we have
\begin{eqnarray}
\frac{\D \xi}{\D r}=\frac{(1-\xi)rA(r)}{r^2+a^2}
\end{eqnarray}
with $A(r):=3+a^2/r^2$. Now the horizon is located at $\xi=0$,
while the asymptotic infinity corresponds to $\xi=1$.
Near the horizon, the radial equation~(\ref{radial_eq}) reduces to
\begin{eqnarray}
&&\xi(1-\xi)\frac{\D ^2R}{\D \xi^2}+(1-\gamma\xi)\frac{\D R}{\D \xi}
\nonumber\\&&\qquad
+\left[\frac{K_*^2}{\xi(1-\xi)A_*^2}-\frac{\Lambda_*(1+a_*^2)}{(1-\xi)A_*^2}\right]R=0,
\end{eqnarray}
where
\begin{eqnarray*}
&&A_*^2:=3+a^2_*,\quad K_*:=(1+a_*^2)\omega_*-a_*m,
\\
&&\gamma:=1-4a_*^2/A_*^2,\quad \Lambda_*:=\lambda+\nu(\nu+1)a_*^2.
\end{eqnarray*}
Performing the transformation $R=\xi^{\alpha}(1-\xi)^{\beta}F(\xi)$
with
\begin{align}
\alpha&=-i\frac{K_*}{A_*},
\\
\beta&=1-\frac{\gamma}{2}-\sqrt{\left(1-\frac{\gamma}{2}\right)^2
-\frac{K_*^2-\Lambda_*(1+a_*^2)}{A_*^2}},
\end{align}
we obtain a hypergeometric differential equation
\begin{eqnarray}
\xi(1-\xi)\frac{\D ^2F}{\D \xi^2}+\left[\tilde c-(1+
\tilde a+\tilde b)\xi\right]\frac{\D F}
{\D \xi}-\tilde a\tilde bF=0,
\end{eqnarray}
where $\tilde a:=\alpha+\beta+\gamma-1$, 
$\tilde b:=\alpha+\beta$, and $\tilde c=1+2\alpha$.
Thus, the near-horizon solution is given in terms of the hypergeometric function by
\begin{align}
R=&A_-\xi^{\alpha}(1-\xi)^{\beta}F(\tilde a,\tilde b,\tilde c;\xi)
\\&+
A_+\xi^{-\alpha}(1-\xi)^{\beta}F(\tilde a-\tilde c+1, \tilde b-
\tilde c+1, 2-\tilde c;\xi),\nonumber
\end{align}
where $A_{\pm}$ are integration constants.
In the limit $r\to r_h$ ($\xi\to 0$) we have
\begin{eqnarray}
R\simeq A_- \xi^{\alpha}+A_+\xi^{\alpha} = A_-e^{-iky}+A_+e^{iky},
\end{eqnarray}
where $y$ is defined earlier in the main text. Since we are imposing the boundary condition
such that no outgoing wave is present at the horizon, we set $A_+=0$.
Thus we arrive at
\begin{eqnarray}
R_{{\rm NH}}=A_-\xi^{\alpha}(1-\xi)^{\beta}
F(\tilde a, \tilde b, \tilde c;\xi)\quad {\rm at}\quad r\simeq r_h.
\label{NH_sol}
\end{eqnarray}
One can check that the convergence condition for the hypergeometric function,
Re$[\tilde c-\tilde a-\tilde b]>0$, is indeed satisfied.

\begin{figure}[tb]
  \begin{center}
    \includegraphics[keepaspectratio=true,height=45mm]{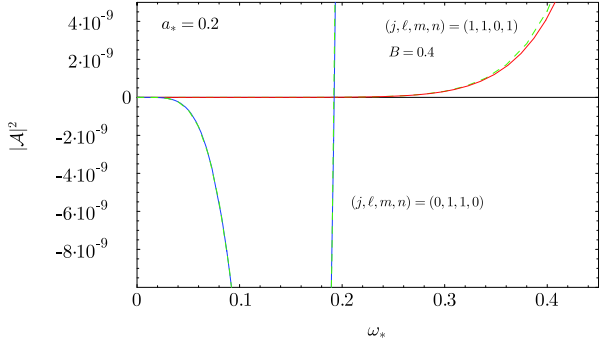}
  \end{center}
  \caption{Absorption probabilities computed analytically and numerically.
  Solid lines follow from the analytic approximation,
  while dashed lines denote our numerical result.}%
  \label{fig:app.eps}
\end{figure}

Now we extend the solution~(\ref{NH_sol}) to go beyond the near-horizon zone.
Using the formula
\begin{eqnarray*}
&&F(\tilde a, \tilde b, \tilde c;\xi) 
= \frac{\Gamma(\tilde c)\Gamma(\tilde c-\tilde a-\tilde b)}
{\Gamma(\tilde c-\tilde a)\Gamma(\tilde c-\tilde b)}
\nonumber\\&&\qquad\qquad\qquad
\times F(\tilde a, \tilde b,\tilde a+\tilde b-\tilde c+1;1-\xi)
\nonumber\\&&\;\;
+(1-\xi)^{\tilde c-\tilde a-\tilde b}\frac{
\Gamma(\tilde c)\Gamma(\tilde a+\tilde b-\tilde c)}
{\Gamma(\tilde a)\Gamma(\tilde b)}
\nonumber\\&&\qquad\qquad\qquad
\times F(\tilde c-\tilde a, \tilde c-\tilde b,\tilde c-
\tilde a-\tilde b+1;1-\xi),
\end{eqnarray*}
and taking the limit $r\to\infty$ ($\xi\to1$) we obtain
\begin{eqnarray}
R_{{\rm NH}}\simeq A_1r^{-3\beta}+A_2r^{-3(2-\gamma-\beta)},\label{NH-far}
\end{eqnarray}
where
\begin{eqnarray}
&&A_1:=A_-(1+a_*^2)^{\beta}r_h^{3\beta}\frac{\Gamma(\tilde c)
\Gamma(\tilde c-\tilde a-\tilde b)}{\Gamma(\tilde c-\tilde a)
\Gamma(\tilde c-\tilde b)},
\nonumber\\
&&A_2:=A_-(1+a_*^2)^{2-\gamma-\beta}r_h^{3(2-\gamma-\beta)}
\frac{\Gamma(\tilde c)\Gamma(\tilde a+\tilde b-\tilde c)}
{\Gamma(\tilde a)\Gamma(\tilde b)}.
\nonumber\\
\end{eqnarray}
The expression~(\ref{NH-far}) should be matched to the far-field solution
which will be derived below.

Let us go on to the far-field solution.
In the far-field zone ($r\gg r_h$) the radial equation~(\ref{radial_eq}) reduces to
\begin{eqnarray}
\frac{\D ^2R}{\D r^2}+\frac{4}{r}\frac{\D R}{\D r}+\left(\omega^2-\frac{\eta}{r^2}\right)R=0.
\end{eqnarray}
The solution is given by
\begin{eqnarray}
R_{{\rm FF}}=B_1\frac{J_{\tau}(\omega r)}{r^{3/2}}+B_2\frac{Y_\tau(\omega r)}{r^{3/2}},
\label{solBess}
\end{eqnarray}
where
$J_\tau$ ($Y_\tau$) is the Bessel function of the first (second) kind,
$\tau:=\sqrt{\eta+9/4}$, and $B_{1, 2}$ are the integration constants.
This solution is in turn to be extended to the near-horizon zone.
Taking the limit $\omega r\ll 1$ we get
\begin{eqnarray}
R_{{\rm FF}}\simeq\frac{B_1}{r^{3/2}\Gamma(\tau+1)}\left(\frac{\omega r}{2}\right)^{\tau}
-\frac{B_2\Gamma(\tau)}{r^{3/2}\pi}\left(\frac{\omega r}{2}\right)^{-\tau}.
\label{FF-near}
\end{eqnarray}

The two solutions~(\ref{NH-far}) and~(\ref{FF-near}) appear to have different powers in $r$,
but by taking the low-energy ($\omega_*\ll1$) and slow-rotation ($a_*\ll 1$) limit
we are able to match them.
Neglecting terms of order $\omega^2_*$, $a_*^2$, and $a_*\omega_*$, we have
$A_*\simeq3$, $\gamma\simeq1$, and $\beta\simeq-(\ell+b_n)/3$,
leading to $-3\beta\simeq\ell+b_n$ and $-3(2-\gamma-\beta)\simeq-(\ell+b_n+3)$
in Eq.~(\ref{NH-far}),
where we used $\eta=(\ell+b_n)(\ell+b_n+3)+{\cal O}(a_*^2\omega_*^2)$.
As for Eq.~(\ref{FF-near}), we have $\tau-3/2\simeq\ell+b_n$ and $-(\tau+3/2)\simeq-(\ell+b_n+3)$.
Thus in this limit we achieve exact matching.

The absorption probability can be expressed in terms of the ratio of the
coefficients $B_r:=B_1/B_2$.
One finds
\begin{align}
B_r=&-\frac{1}{\pi}\left[\frac{\omega_*(1+a_*^2)^{1/3}}{2}\right]^{-(2\ell+2b_n+3)}
\label{B_r}
\nonumber\\&\times
\frac{\tau\Gamma^2(\tau)\Gamma(\tilde a)\Gamma(\tilde b)
\Gamma(\tilde c-\tilde a-\tilde b)}
{\Gamma(\tilde c-\tilde a)\Gamma(\tilde c-\tilde b)
\Gamma(\tilde a+\tilde b-\tilde c)}.
\end{align}
In the limit $r\to\infty$ the solution~(\ref{solBess}) can be written as
\begin{eqnarray}
R_{{\rm FF}}\simeq A_{\rin}^{(\infty)}\frac{e^{-i\omega r}}{r^2}
+A_{\rout}^{(\infty)}\frac{e^{i\omega r}}{r^2},
\end{eqnarray}
where
\begin{eqnarray}
A_{\rin}^{(\infty)}&=&(B_1+iB_2)\frac{e^{i(\tau\pi/2+\pi/4)}}{\sqrt{2\pi\omega}}
\\
A_{\rout}^{(\infty)}&=&(B_1-iB_2)\frac{e^{-i(\tau\pi/2+\pi/4)}}{\sqrt{2\pi\omega}}.
\end{eqnarray}
The absorption probability is given by 
$|{\cal A}|^2=1-|A_{\rout}^{(\infty)}/A_{\rin}^{(\infty)}|^2$.
In the low energy limit ($\omega_*\ll 1$) we have $B_rB_r^*\gg(B_r-B^*_r)/i\gg1$.
Hence,
\begin{eqnarray}
|{\cal A}|^2\simeq2i\left(\frac{1}{B_r}-\frac{1}{B_r^*}\right).\label{ABB*}
\end{eqnarray}
Substituting~(\ref{B_r}) to~(\ref{ABB*}) results in a lengthy expression, but
when $m$ is not large we may expand the equation with respect to $\alpha\,(\ll1)$
in the low-energy and slow-rotation regime.
Then, at leading order in $\alpha=-iK_*/A_*$, one finds
\begin{widetext}
\begin{eqnarray}
|{\cal A}|^2=\frac{4\pi K_*}{A_*}
\frac{\Gamma^2(2\beta+\gamma-2)\Gamma^2(1-\beta)(2-2\beta-\gamma)\sin^2[\pi(2\beta+\gamma)]}
{(\ell+b_n+3/2)\Gamma^2(\ell+b_n+3/2)\Gamma^2(\beta+\gamma-1)\sin^2[\pi(\beta+\gamma)]}
\left[\frac{\omega_*(1+a_*^2)^{1/3}}{2}\right]^{2\ell+2b_n+3}.
\label{absorption2}
\end{eqnarray}
\end{widetext}
In Fig.~\ref{fig:app.eps} we compare this analytic result with our numerical calculation
employed in the main text. It can be seen that the numerical result indeed agrees with
the analytic one in the low-energy and slow-rotation regime.
Since all the terms except $K_*$ on the right hand side of Eq.~(\ref{absorption2})
are positive, the sign of $|\mathcal A|^2$ is controlled by
$K_*=(1+a_*^2)\omega _*-a_*m=r_h(1+a_*^2)(\omega -m\Omega_h )$.
Therefore, the absorption probability is negative for superradiant
modes ($0<\omega <m\Omega_h $).

In the low energy regime the $j=\ell=m=n=0$ mode will be dominant,
for which the absorption probability is given by
\begin{align}
|\mathcal A_{0000} |^2 = \frac{4 \omega_*^4(1+a_*^2)^2}{3(3+a_*^2)(2-\gamma )}+\cdots.
\end{align}
This does not depend on the deficit angle.

To obtain the expression for the absorption cross section we extract the ingoing s-wave
from the plane wave~\cite{Das,Das2}:
\begin{align}
e^{i\omega z} \to \frac{\mathcal K}{(A_4B)^{1/2}r^2}e^{-i\omega r}
+(\textrm{higher multipole moments}),
\label{plane}
\end{align}
where the square root in the denominator is the normalization factor.
In order to determine $\mathcal K$, one integrates both sides of
Eq.~(\ref{plane}) over the 4-sphere with a deficit angle, and then, 
looking at the far region $\omega r\gg 1$, extracts only the ingoing modes.
Thus, we arrive at
\begin{align}
 |\mathcal K|^2 =\frac{2^4(A_3)^2\Gamma^2 (2) B}{4\omega^4 A_4}=\frac{6\pi^2
 B}{\omega ^4},
\end{align}
leading to the low energy absorption cross section
\begin{align}
\sigma _0&=|\mathcal A|^2 |\mathcal K|^2 \nonumber \\&=\left[
\frac{8\pi ^2}{3}r_{ h}^2 (r_{h}^2+a^2)B\right]\frac{(1+a_*^2)}
{(1+a_*^2/3)(2-\gamma )}+\cdots .
\label{crosssection}
\end{align} 
In the static limit we have $\sigma _0 =$ (horizon area), 
which reproduces the general result for the spherically symmetric case~\cite{Das}.



\end{document}